\documentstyle[aps,prl,floats,graphicx,epsfig]{revtex}
\begin{document}
\draft

\twocolumn[\hsize\textwidth\columnwidth\hsize\csname @twocolumnfalse\endcsname

\title{Critical velocity and event horizon in pair-correlated systems with
"relativistic" fermionic quasiparticles.}
\author{N.B. Kopnin$^{1,2}$ and G.E. Volovik$^{1,2}$}
\address{
$^1$Helsinki University of Technology, Low Temperature Laboratory,
P. O. Box 2200, FIN-02015 HUT, Finland\\
$^2$Landau Institute for Theoretical Physics, 117334 Moscow,
Russia
}

\date{\today}
\maketitle

\begin{abstract}
The condition for the appearance of event horizon is considered in such
pair-correlated systems (superfluids and superconductors) where the fermionic
quasiparticles obey the  "relativistic" equations. In these systems, the
Landau
critical velocity of superflow  corresponds to the speed of light. In
conventional systems, such as $s$-wave superconductors, the superflow remains
stable even above the Landau treshold. We showed that in the "relativistic"
systems however the quantum vacuum becomes unstable and  the superflow
collapses
after  the "speed of light" is reached, so that horizon cannot appear. Thus an
equilibrium dissipationless superfluid flow state and the horizon are
incompatible due to quantum effects. This negative result is consistent
with the
quantum Hawking radiation from the horizon, which would lead to the dissipation
of the flow.

\end{abstract}

\

\pacs{PACS numbers:     }

\

] \narrowtext
%\twocolumn

\section{Introduction}

It is known that the some aspects of the problem of the black holes can be
modelled in condensed matter physics
\cite{UnruhSonic,Jacobson1991,Visser1997,JacobsonVolovik,RotatingCore}. This
comes from the fact that the acoustic waves propagating in the moving
classical liquid
\cite{UnruhSonic,Jacobson1991,Visser1997} and the fermions propagating in the
texture of superfluid $^3$He-A
\cite{JacobsonVolovik,Exotic,Volovik1996} obey the
relativistic-type equations in curved space, which metric is produced by the
flow field and in the  $^3$He-A also by texture. In both systems the
corresponding velocity of light  can be exceeded, which leads to the
possibility to investigate the event horizon problem.

The  $^3$He-A, as well as other pair-correlated systems (including the
$d$-wave superconductors, which also contains the relativistic fermions, see eg
\cite{SimonLee}), are better model for simulations of the event horizon
than the
classical liquid, since these are the quantum systems, which ground states are
in many respects similar to the quantum vacuum of high-energy physics. That is
why they can be used for the investigation of the quantum effects related with
the event horizon, such as Hawking radiation \cite{Hawking} and statistical
entropy \cite{Bekenstein}.

Here we address the stability problem of the quantum vacuum in the presence of
the event horizon: whether a nondissipative flow of superfluid is possible
in the presence of horizon, or the horizon always leads to a vacuum
reconstruction into a state with dissipation. This can be considered using an
example of a superflow with velocity  exceeding the Landau critical
velocity $v_L$.

Let the superfluid move at $T=0$ with so-called superfluid
velocity
${\bf v}_s$, while the walls of container, the distinguished reference frame,
move with the so-called normal velocity
${\bf v}_n$. The physical properties of the vacuum state depend
on the relative (counterflow) velocity
${\bf w}={\bf v}_s-{\bf v}_n$. In the subcritical regime, $w<v_L$, the order
parameter is independent of $w$: the observer moving with
${\bf v}_s$ does not see any difference in the liquid as compared to the case
when  ${\bf v}_s={\bf v}_n$. The system thus retains some kind of Galilean
invariance even in the presence of the container wall. For example the
density of
superfluid component in the expression for the mass current ${\bf j}=\rho_s{\bf
v}_s +\rho_n{\bf v}_n$, which represents the vacuum component of the liquid,
does not depend on
$w$ and coincides with the total density:
$\rho_s(T=0,w<v_L)=\rho$, while the density of the normal component, which
represents the matter, is always zero:
$\rho_s(T=0,w<v_L)=\rho-\rho_s(T=0,w<v_L)=0$.

The observer starts to see the dependence on the velocity with respect to the
reference frame if $w$ exceeds the Landau critical velocity $v_{L}$,
at which the negative energy levels appear, ie the states with the
negative doppler shifted energy $E_{\bf p} + {\bf p}\cdot {\bf w}<0$. If the
system is fermionic, then the typical situation in the supercritical flow
regime
above $v_L$ is the following: The negative energy levels become finally
occupied
and  after such vacuum reconstruction the system obtains a new
equilibrium state again with the frictionless superflow. However all the
physical quantities  do now  depend on
$w$, as will be seen by comoving observer. The vacuum state becomes
anisotropic,
the superfluid density does depend  on $w$ and becomes less than the total mass
density,
$\rho_s(T=0,w>v_L)<\rho$. The other part of the liquid with the so-called
normal
density $\rho_n=\rho-\rho_s$ comprises the normal component (matter) and
consists of the trapped fermions with the negative energy. In
equilibrium  this component is at rest  with respect to the container
reference frame, ie it  moves with the velocity
${\bf v}_n$.  The previous
"Galilean" symmetry is thus broken by created matter.

There is another critical velocity, $v_c$, at which the superfluid vacuum
is exhausted, ie the superfluid density completely disappears,
$\rho_s(w=v_c)=0$, and thus the nondissipative superflow does not exist
anymore.
Typically $v_c>v_L$ and thus the violation of the Galilean symmetry occurs
earlier than the superfluidity collapses. This happens for example in
conventional $s$-wave superconductors, where $v_L$ and $v_c$ are of
the same order with $v_c>v_L$,  and also in superfluid
$^3$He-A, where $v_L=0$, while $v_c$ is finite (see eg
\cite{Vollhardt1990,Muzikar1983,Nagai1984}).

Existence of region with the stable supercritical superflow, $v_L<w<v_c$,
allows
us to ask the question concerning the quantum effect of the event horizon. Let
us consider the system, where quasiparticles are described by the
effective  relativistic equations, such that the Landau velocity corresponds to
the "speed of light"
$c$. In this case the supercritical velocity $w>v_L$ corresponds to the
superluminal one and thus the event horizon can be constructed (see below).
The problem appears if one takes into account the quantum effects: on one hand
the frictionless superflow in the regime  $v_L<w<v_c$ is stable, and this
stability cannot be violated by the presence of horizon, on the other
hand the Hawking radiation from horizon means that such superflow is
dissipative
in the presence of  horizon.

Thus we have a dilemma: (i) either one should doubt the fundamentality
of the Hawking radiation from the event horizon in the supercritical regime,
(ii) or in the relativistic-type systems the "superluminal"  regime of
superflow
is prohibited, ie $v_L=v_c=c$. Here we consider the pair-correlated fermionic
system with the superconducting/superfluid state of the polar-type, which has
the "relativistic" Bogoliubov fermionic qiasiparticles. We find that in this
system the second alternative occurs: the nondissipative superflow collapses at
$w=v_L=c$, which means that the horizon never appears in the stationary
nondissipative superflow: it can exist only  in the dissipative flow state.

\section{Gap equation for "speed of light".}

The energy spectrum of the pair-correlated system and its vacuum state are
determined by the self-consistent equations for the so called gap function
$\Delta_{\bf p}$ which determines the quasiparticle spectrum and thus the
"speed
of light" $c$:
\begin{equation}
\Delta_{\bf p}= \sum_{\bf p'}V_{ {\bf p},{\bf p}'} {\Delta_{{\bf p}'}\over
E_{{\bf p}'}}\left(1-n_{{\bf p}'}-n_{-{\bf p}'}\right) ~.
\label{GapEquation}
\end{equation}
Here $V_{ {\bf p},{\bf p}'}$ is the pairing potential, $E_{\bf p}$ is the
quasiparticle energy in the pair-correlated state and $n_{{\bf p}}$ is their
thermal distribution
\begin{equation}
E_{\bf p}=\sqrt{\Delta^2_{\bf p} +\epsilon^2_{\bf p}}~~, ~~n_{{\bf p}}={1\over
1+\exp {E_{\bf p}+{\bf p}\cdot {\bf w} \over T} }
\label{Eandn}
\end{equation}
 $\epsilon_{\bf p}=(p^2-p_F^2)/2m$ is the fermion energy in the absence of
the pair correlation. If the superflow velocity ${\bf v}_s$ deviates from the
container reference frame velocity ${\bf v}_s$, the distribution function
is Doppler shifted; further we assume that
${\bf v}_n=0$ and thus ${\bf w}={\bf v}_s$.

Let us consider how the vacuum state ($T=0$) is disturbed by the counterflow
$w$ in the supercritical regime. We are interested in the case when the
spectrum
of quasiparticles is "relativistic", so that the horizon problem can arise. For
this reason we consider the two-dimensional case, spin-triplet pairing with the
orbital momentum $L=1$ for which the pairing potential $V_{ {\bf p},{\bf p}'}=
2(V_1/p_F^2){\bf p}\cdot {\bf p'}$, and the gap function corresponding to the
polar phase:\cite{Vollhardt1990}
\begin{equation}
\Delta_{\bf p}= cp_x~,
\label{PolarPhase}
\end{equation}
where the factor $c$ plays the part of the speed of light.

Let the velocity of the superflow be
along the same axis
$x$, ie
${\bf v}_s=w\hat x$, thus the superflow does not violate the symmetry of the
polar state and the Eq.(\ref{PolarPhase}) remains the solution even in the
presence of the superflow. The Doppler shifted energy of the fermions in this
pair-correlated state
\begin{equation}
E(p_x,\epsilon)=E_{\bf p}+{\bf p}\cdot {\bf v}_s=\sqrt{\epsilon^2  + c^2p_x^2}+
wp_x
\label{RelativisticSpectrum1}
\end{equation}
or
\begin{equation}
(E-wp_x)^2= \epsilon^2  + c^2p_x^2
\label{RelativisticSpectrum2}
\end{equation}
This corresponds to the relativistic 1D particle with the mass $\epsilon$
moving in the Lorentzian metric:
\begin{equation}
g^{00}=-1~~,~~g^{01}=w~~,~~g^{11}=c^2-w^2~~.
\label{metric}
\end{equation}
If the counterflow velocity $w$ can exceed $c$, then one can construct the
inhomogeneous flow state with such coordinate dependence of $w(x)$ and $c(x)$,
that $w(x)$ crosses $c(x)$. In this case the metric element $g^{11}(x)=c^2-w^2$
crosses zero at the points
$x=x_{h}$ where
$w(x_h)=c(x_h)$ and thus the event horizon appears at these points.

From
Eq.(\ref{GapEquation}) one   obtains  that if
$w<c$, then at
$T=0$ there is no quasiparticles: $n_{\bf p}=0$, ie the vacuum remains intact.
The system is effectively Galilean invariant and the speed of light is
independent of $w$.

The problem is whether the velocity of the flow $w$ can exceed the Landau
velocity $v_L$ which is now the "speed of light" $c$. If yes, then the horizon
can be constructed. Let us consider the gap equation
Eq.(\ref{GapEquation}) in the case when $w>c$. We assume that the speed
of light is small compared to the Fermi velocity, $c\ll v_F=p_F/m$, which is
typical for the weakly interacting Fermi-liquid. In this case the momentum is
concentrated near the Fermi-momentum,
${\bf p}=(p_F\sin\phi,p_F\cos\phi)$, and one can write
\begin{equation}
 \sum_{\bf p} =\int  {d^2p\over (2\pi)^2} ={m\over 2\pi}\int d\epsilon
\int{d\phi\over 2\pi}
\label{SumAsIntegral}
\end{equation}

In principle, one can expect that at $w>c$ the spead of light becomes
dependent on $w$. Thus let us introduce the bare speed of light
$c_0=c(w=0)$ and the current speed of light $c(w)$ if $w>c(w)$.
As we have seen, $c(w<c)=c_0$, and this solution persists until $w$ reaches the
Landau velocity
$c_0$. Thus the first branch of $c(w)$ is
\begin{equation}
c_1(w)=c_0~~,~~w\leq c_0~~.
\label{FirstBranch-c(w)}
\end{equation}

\section{State with  "superluminal" flow.}

If
$w>c$, the Galilean invariance becomes broken due to fermions filling the
negative levels of the energy in Eq.(\ref{RelativisticSpectrum1}). The
number of
particles on the negative energy levels is the fermionic step function of the
energy
\begin{equation} n_{{\bf p}}=\Theta\left(-E(p_x,\epsilon)\right)~.
\label{FillingLevels}
\end{equation}

Then from Eq.(\ref{GapEquation})  one obtains the following
equation for the factor
$c(w)$ in the gap function Eq.(\ref{PolarPhase}):
\begin{eqnarray}
\nonumber\int_0^\infty d\epsilon
\int_0^{2\pi}{d\phi\over 2\pi}\left( {\sin^2\phi\over
\sqrt{\epsilon^2+c^2\sin^2\phi} } -  {\sin^2\phi\over
\sqrt{\epsilon^2+c_0^2\sin^2\phi} }\right)\\
=  \int_0^\pi{d\phi\over
2\pi} \sin^2\phi \int_0^{\sin\phi\sqrt{w^2-c^2}} { d\epsilon\over
\sqrt{\epsilon^2+c^2\sin^2\phi}}
\label{c(w)equation}
\end{eqnarray}
Note that the details of pair interaction are concealed in the bare speed of
light $c_0$, determined by this interaction. From Eq.(\ref{c(w)equation}) one
has
\begin{equation}
\ln{c_0\over c}=\rm{arsh}\left(\sqrt{{w^2\over c^2}-1}\right)
\label{SecondBranchEquation}
\end{equation}
which gives the solution for the "speed of light" $c$ in the superluminal
regime
$w>c(w)$:
\begin{equation}
c_2(w) =c_0\sqrt{{2w\over
c_0}-1}~~, ~~{1\over 2} c_0<w<c_0~.
\label{SecondBranch-c(w)}
\end{equation}
It follows that no solution exists above the Landau velocity, ie at $w>c_0$,
which means that the Landau velocity coincides with the velocity of the
superflow collapse and thus with the bare speed of light: $v_L=v_c=c_0$.

Below the Landau velocity one has two branches, $c_1(w)=c_0$ and $c_2(w)$ (see
Figure). Both can be obtained as etxrema of the superfluid vacuum energy in the
presence of the mass current
\begin{eqnarray}
\nonumber
\Omega_S(s,w)-\Omega_N=-jw +{1\over 2}\rho w^2 -{1\over 4}\rho c^2 -{1\over
2}\rho c^2
\ln{c_0\over c}
\\
+ {1\over 2}\rho c^2 \left\{\ln\left[{w\over c} +\sqrt{{w^2\over
c^2}-1}\right]- {w\over c}\sqrt{{w^2\over
c^2}-1}\right\}\Theta(w-c)
\label{GeneralEnergy}
\end{eqnarray}
Here $\Omega_N$ is the free energy of the normal state, ie at $c=0$; the
mass density in this 2D model is  $\rho=mp_F^2/2\pi \hbar^2$;  the first
term $-jw$ means that the free energy is to be extremized at the given
mass current energy ${\bf j}=\rho {\bf v}_s
+ \sum_{\bf k} {\bf k} n_{\bf k}$.
The current in a given state  (see
Figure) can be obtained from the extremum of the vacuum
energy with respect to $w$:
$\partial\Omega_S/\partial w=0$. This gives the general expression for the mass
current density
\begin{equation}
j(w,c) =\rho \left(w-\Theta(w-c)\sqrt{ w^2- c^2}\right)~.
\label{CurrentGeneral}
\end{equation}

%%%%%%%%%%%%%%%%%%%%%%%%%%%%%%%%%%%%%%%%%%%%%%%%%%%%%%%%%%%
\begin{figure}[!!!t]
%\centerline{\epsfxsize=0.40\textwidth\epsfbox{Figc(w).eps}}
%\bigskip
\begin{center}
\leavevmode
\epsfig{file=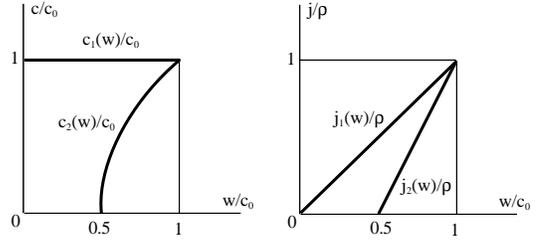,width=0.8\linewidth}
\caption[c(w)]
    {Two states of the supeflow. In the subluminal flow state the
speed of light $c_1(w)=c_0$ does not depend on flow velocity $w$ with
respect to the container wall. The superluminal state with $c_2(w)<w$ is
locally unstable.  The mass current in these two states is presented on the
right.}
\label{c(w)}
\end{center}
\end{figure}
%%%%%%%%%%%%%%%%%%%%%%%%%%%%%%%%%%%%%%%%%%%%%%%%%%%%%%%%%%

The second branch,  corresponding to the superluminal flow
$c_2(w)<w$, represents the saddle point solution of the vacuum
energy and  thus is unstable towards the formation of
the regular branch,  corresponding to the subluminal flow,
$c_1(w)=c_0<w$. This second branch with similar behavior has been also
found for
the $^3$He-B under the superflow \cite{Vollhardt1980}.

\section{Conclusion}

We found that in the  superfluid  analogs of the relativisic system, the stable
superflow with velocity exceeding the corresponding "speed of light",
$w>c$, does
not exist and thus the dissipationless state with horizon does not appear.  The
collapse of the superfluid quantum vacuum in the superluminal regime is
compatible with the Hawking radiation, which leads to the dissipation in
the presence of horizon and thus cannot exist in the stable superflow. Horizon
can appear only if the flow state is dissipative. This can happen if the
external body or the order parameter texture moves in the superfluid with the
supercritical velocity, as was discussed in
\cite{JacobsonVolovik} for the case of moving topological soliton in $^3$He-A.
The Hawking radiation gives rise to the additional dissipation during the
motion
of the object.

\acknowledgements
GEV thanks Ted Jacobson for illuminating discussions. This work was
supported by
the Russian Foundation for Fundamental Research grant No. 96-02-16072, by
the RAS
program ``Statistical Physics'', by the Intas grant 96-0610 and by European
Science Foundation.

\end{document}